\documentclass[pra,twocolumn,superscriptaddress,showpacs]{revtex4-2}

\usepackage{amsmath,amssymb}
\usepackage{graphicx}
\usepackage{bm}
\usepackage{hyperref}
\hypersetup{pdftitle={Amplitude Encoding of Slater-Type Orbitals via Matrix Product States: Efficient State Preparation and Integral Evaluation on Quantum Hardware}}
\usepackage{braket}
\usepackage{booktabs}
\usepackage{multirow}
\usepackage{xcolor}

\begin{document}

\title{Amplitude Encoding of Slater-Type Orbitals via Matrix Product States:\texorpdfstring{\\}{} Efficient State Preparation and Integral Evaluation on Quantum Hardware}

\author{Sorin Bolos}
\email{sorin.bolos@transilvania-quantum.com}
\affiliation{Transilvania Quantum}
\affiliation{Technical University of Cluj-Napoca, Romania}

% \author{Prof. Tudor Palade}
% \affiliation{Technical University of Cluj-Napoca, Romania}

\date{\today}

\begin{abstract}
Slater-type orbitals (STOs) provide the physically correct description of atomic wavefunctions but have been largely replaced by Gaussian-type orbitals in computational chemistry due to the lack of closed-form multi-center integrals. We present a systematic study of amplitude encoding of STOs on quantum computers using matrix product states (MPS). For one-dimensional orbital functions of the form $p_d(x) e^{-\zeta x}$, we derive analytical MPS constructions with constant bond dimension $\chi = d + 1$, requiring $O(n)$ classical and quantum resources for $n$-qubit registers with no grid sampling. We demonstrate a complete one-electron integral pipeline---overlap, kinetic energy, and nuclear attraction---in one dimension, validating the overlap and kinetic energy on IBM Heron processors at 5~qubits with 0.67\% hardware-induced error using Zero-Noise Extrapolation. In three dimensions, we compute multi-center overlap integrals between 1s and 2s orbitals in Cartesian coordinates with 0.02\% discretization error at 18~qubits. A systematic entanglement analysis reveals that the MPS bond dimension of three-dimensional STOs in Cartesian coordinates saturates with increasing grid resolution---reaching $\sim$138 for the hydrogen 1s orbital at 12~qubits per coordinate---establishing bounded encoding complexity rather than the exponential scaling initially expected. The SVD truncation threshold provides a practical resource parameter, reducing the bond dimension to 39 at threshold $10^{-6}$ with negligible accuracy loss. These results map the entanglement landscape for amplitude encoding of atomic orbitals and establish MPS-based state preparation as a viable path toward exact STO basis sets on quantum computers.
\end{abstract}

\maketitle

%% ============================================================
%% SECTION I: INTRODUCTION
%% ============================================================

\section{Introduction}
\label{sec:introduction}

The choice of basis set is among the most consequential approximations in computational quantum chemistry. Slater-type orbitals (STOs), which take the form $r^{n-1}e^{-\zeta r}Y_{lm}(\theta,\varphi)$, provide the physically correct description of atomic wavefunctions: they satisfy the nuclear cusp condition at $r=0$, exhibit the correct exponential decay at large $r$, and form a complete basis for the hydrogen atom~\cite{Slater1930}. Despite these advantages, STOs have been largely abandoned in practical molecular calculations in favor of Gaussian-type orbitals (GTOs), which take the form $r^{n-1}e^{-\alpha r^2}Y_{lm}(\theta,\varphi)$. The reason is purely computational: the Gaussian product theorem allows multi-center integrals over GTOs to be evaluated in closed form, while the corresponding integrals over STOs lack closed-form solutions and require expensive numerical methods~\cite{HehreStewartPople1969}.

This trade-off introduces a systematic approximation. To compensate for the incorrect behavior of individual Gaussians---the wrong cusp at the nucleus, the too-rapid decay at large distance---practical calculations use contracted basis sets such as STO-$n$G, in which each STO is approximated by a linear combination of $n$ Gaussians~\cite{SzaboOstlund}. While effective, this approximation introduces basis set error that propagates into all subsequent calculations, including correlated methods such as configuration interaction and coupled cluster. The error is systematic and well-characterized but fundamentally unavoidable within the Gaussian framework.

Quantum computers offer an alternative path. The amplitude encoding paradigm stores a function $f(x)$ evaluated on $2^n$ grid points as the amplitudes of an $n$-qubit quantum state, achieving an exponential compression of the classical representation. If the state can be prepared efficiently---in $O(\text{poly}(n))$ quantum gates---then inner products between amplitude-encoded states provide discretized integrals at a cost independent of the grid size. This suggests a route to exact STO integrals on quantum hardware: encode the orbital directly as a quantum state, without the Gaussian approximation, and extract the integral from the quantum inner product.

The feasibility of this approach depends on a single question: can the target orbital be prepared efficiently? Generic quantum states of $n$ qubits require $O(2^n)$ gates to prepare~\cite{Shende2006}, negating any advantage over classical computation. Efficient preparation is possible only when the state has limited entanglement---specifically, when its matrix product state (MPS) representation has bounded bond dimension. The bond dimension $\chi$ quantifies the entanglement across bipartitions of the qubit register and directly determines the circuit depth: each MPS tensor becomes a unitary gate acting on $1 + \lceil\log_2 \chi\rceil$ qubits, and $n$ such gates are applied sequentially. Constant $\chi$ gives $O(n)$ total gates---exponentially cheaper than the $2^n$ grid points being encoded.

Several approaches to quantum state preparation have been explored in the literature, including variational methods~\cite{Peruzzo2014}, quantum read-only memory (QROM)~\cite{Babbush2018}, and arithmetic circuits~\cite{Haner2018}. Variational methods suffer from barren plateaus in the optimization landscape when the ansatz is sufficiently expressive~\cite{McClean2018}. QROM achieves exact preparation but at $O(2^n)$ cost, offering no advantage over classical computation. MPS-based preparation~\cite{Schon2005,Ran2020} occupies a middle ground: it is constructive (not variational), exact (not approximate), and efficient when the bond dimension is bounded. The approach has been applied to various quantum simulation tasks, but a systematic study of its application to atomic orbital encoding---including the characterization of which orbitals admit efficient preparation and which do not---has not been undertaken.

In this work, we address this gap. We present a systematic study of MPS-based amplitude encoding of Slater-type orbitals, spanning one-dimensional and three-dimensional representations across spherical and Cartesian coordinate systems. Our contributions are:

\begin{enumerate}
\item[(i)] Analytical MPS constructions for one-dimensional orbital functions of the form $p_d(x) e^{-\zeta x}$ where $p_d$ is a polynomial of degree $d$, achieving constant bond dimension $\chi = d+1$ with $O(n)$ classical and quantum resources. The MPS tensors are derived from the orbital parameters alone, requiring no grid sampling.

\item[(ii)] \begin{sloppypar}A complete one-electron integral evaluation pipeline---overlap, kinetic energy, and nuclear attraction---demonstrated in one dimension with statevector simulation and validated on IBM Heron quantum processors. The overlap integral at 5~qubits achieves 0.67\% hardware-induced error with Zero-Noise Extrapolation.\end{sloppypar}

\item[(iii)] Extension to three dimensions: multi-center overlap integrals between 1s and 2s orbitals computed in Cartesian coordinates with 0.02\% discretization error at 6~qubits per coordinate (18 total qubits), and same-center orthogonality verified in both spherical and Cartesian coordinates.

\item[(iv)] A systematic entanglement analysis revealing that the bond dimension of three-dimensional STOs in Cartesian coordinates saturates with increasing grid resolution, reaching an asymptotic maximum of $\sim$138 for the hydrogen 1s orbital. The cross-coordinate entanglement likewise saturates. This establishes that Cartesian amplitude encoding has bounded complexity---expensive but not intractable---in contrast to initial expectations of exponential scaling.

\item[(v)] Characterization of the SVD truncation threshold as a practical resource parameter: reducing the threshold from $10^{-12}$ to $10^{-6}$ cuts the bond dimension by a factor of 3.4 with negligible impact on integral accuracy.
\end{enumerate}

The remainder of this paper is organized as follows. Section~\ref{sec:theory} develops the theoretical framework: amplitude encoding, MPS decomposition, analytical tensor constructions, and one-electron integral circuits. Section~\ref{sec:results} presents the results: one-dimensional integrals with hardware validation, three-dimensional overlap integrals, and the entanglement analysis across coordinate systems and scaling regimes. Section~\ref{sec:discussion} discusses the implications, limitations, and path to multi-center and two-electron integrals. Section~\ref{sec:conclusion} concludes.

%% ============================================================
%% SECTION II: THEORY
%% ============================================================

\section{Theory}
\label{sec:theory}

\subsection{Amplitude encoding and integral evaluation}
\label{sec:amplitude_encoding}

A function $f(x)$ defined on a discrete grid of $N = 2^n$ points can be stored as the amplitudes of an $n$-qubit quantum state:
\begin{equation}
\ket{f} = \frac{1}{\mathcal{N}} \sum_{j=0}^{N-1} f(x_j) \ket{j},
\end{equation}
where $x_j = j \cdot \Delta x$ are the grid points, $\Delta x = L/N$ is the spacing, $L$ is the total simulated range, and $\mathcal{N} = \sqrt{\sum_j |f(x_j)|^2}$ is the normalization constant. This representation stores $2^n$ function values using $n$ qubits---an exponential compression of the classical data.

The inner product of two amplitude-encoded states gives a discretized integral:
\begin{equation}
\braket{f|g} = \frac{1}{\mathcal{N}_f \mathcal{N}_g} \sum_j f^*(x_j)\, g(x_j),
\end{equation}
which approximates $\int f^*(x) g(x)\, dx$ up to the normalization factors and the grid spacing $\Delta x$. The physical integral is recovered as:
\begin{equation}
\int f^*(x)\, g(x)\, dx \approx \mathcal{N}_f \cdot \mathcal{N}_g \cdot \braket{f|g} \cdot \Delta x.
\label{eq:physical_integral}
\end{equation}
The normalizations $\mathcal{N}_f$, $\mathcal{N}_g$ are classical quantities computed during state preparation. Only the inner product $\braket{f|g}$ is evaluated on the quantum computer.

This extends naturally to higher dimensions. Two registers of $n$ qubits each encode a two-dimensional function on a $2^n \times 2^n$ grid:
\begin{equation}
\ket{h} = \frac{1}{\mathcal{N}} \sum_{j,k} h(x_j, y_k) \ket{j}\ket{k},
\end{equation}
and the inner product $\braket{\Phi|h}$ gives a discretized double integral $\iint \Phi^* h\, dx\, dy$. A single inner product on the tensor product space replaces the double summation. Similarly, three registers encode a 3D function on a $2^n \times 2^n \times 2^n$ grid.

The computational advantage of this approach depends on two factors: the cost of preparing the amplitude-encoded state $\ket{f}$, and the cost of extracting the inner product from measurements. If state preparation requires $O(\text{poly}(n))$ gates --- exponentially fewer than the $2^n$ grid points ---, the quantum representation is efficient. The inner product is extracted via the compute/uncompute method or Hadamard test, requiring $O(1/\varepsilon^2)$ repetitions for precision $\varepsilon$ through sampling, or $O(1/\varepsilon)$ with amplitude estimation~\cite{Brassard2002}.

\subsection{Matrix product state decomposition}
\label{sec:mps}

A state vector of $2^n$ amplitudes can be decomposed as a matrix product state (MPS)~\cite{Schollwock2011}:
\begin{equation}
\ket{\psi} = \sum_{\sigma_1 \ldots \sigma_n} A^{\sigma_1} A^{\sigma_2} \cdots A^{\sigma_n} \ket{\sigma_1 \sigma_2 \ldots \sigma_n},
\end{equation}
where each $A^{\sigma_i}$ is a matrix of dimension $\chi_{i-1} \times \chi_i$, with $\sigma_i \in \{0,1\}$ the physical index and $\chi_i$ the bond dimension at cut~$i$. The boundary conditions are $\chi_0 = \chi_n = 1$, making the product a scalar for each bit string.

The bond dimension $\chi_i$ at cut $i$ equals the Schmidt rank across the bipartition $\{1,\ldots,i\}|\{i{+}1,\ldots,n\}$. It quantifies the entanglement across that cut: $\chi = 1$ means the state is a product state across the cut, while $\chi = 2^{min(i, n-i)}$ indicates maximal entanglement. The maximum bond dimension $\chi_{\max} = \max_i \chi_i$ determines the computational resources required for the MPS circuit. Specifically:

\begin{itemize}
\item \emph{Bond qubits}: the circuit requires $\lceil\log_2 \chi_{\max}\rceil$ ancilla qubits to represent the bond index. A state with $\chi_{\max} = 2$ needs 1 bond qubit; $\chi_{\max} = 39$ needs 6; $\chi_{\max} = 132$ needs 8.
\item \emph{Gate size}: each MPS tensor $A^{\sigma_i}$ is implemented as a unitary gate acting on one physical qubit plus all bond qubits — a gate on $(1 + \lceil\log_2 \chi_{\max}\rceil)$ qubits total.
\item \emph{Transpiled cost}: a unitary on $m$ qubits decomposes into $O(4^m)$ two-qubit gates in the worst case~\cite{Shende2006}, though structured unitaries from MPS tensors are typically cheaper. For $\chi_{\max} = 2$ ($m = 2$), each gate costs $O(1)$ two-qubit gates. For $\chi_{\max}= 39$ ($m = 7$), the cost per gate is substantially higher.
\item \emph{Total circuit}: $n$ sequential gates, giving $O(n)$ depth with a per-gate cost determined by $\chi_{\max}$. The total two-qubit gate count scales as $O(n \times 4^{\lceil\log_2 \chi_{\max}\rceil})$.
\end{itemize}

The bond dimension is therefore the single quantity that determines whether amplitude encoding is practical: constant $\chi$ gives $O(n)$ total gates, polynomially growing $\chi$ gives polynomial total gates, and exponentially growing $\chi$ makes the approach infeasible. This scaling follows from the fact that each entangling gate crossing a bipartition can at most double the Schmidt rank across that cut ~\cite{NielsenChuang}, giving a lower bound of $\log_2(\chi)$ entangling gates per cut.

The MPS tensors are obtained by sequential singular value decomposition (SVD) of the state vector. We use the right-canonical form, in which the SVD proceeds from the least significant bit to the most significant bit. At each step, the state is reshaped into a matrix, the SVD is computed, and the right-singular vectors form the MPS tensor for that site while the singular values are absorbed into the remainder.
 
The right-canonical form is required for the circuit construction: it guarantees that the isometry columns derived from each tensor are orthonormal in the circuit's computational basis, allowing direct completion to a unitary gate via QR decomposition of the orthogonal complement. The left-canonical form (SVD from most significant to least significant bit) produces orthonormal columns in a different basis and fails to yield valid circuit unitaries---a subtle but critical distinction.

Singular values below a chosen threshold are discarded during the SVD, controlling the trade-off between encoding accuracy and bond dimension. We find that a threshold of $10^{-6}$ reduces the bond dimension by up to a factor of 3.4 compared to $10^{-12}$ with negligible impact on computed integrals (Sec.~\ref{sec:entanglement}).

\subsection{Analytical MPS for 1D orbital functions}
\label{sec:analytical_mps}

All constructions in this subsection apply to one-dimensional functions on a single register of $n$ qubits. For several classes of orbital functions, the MPS tensors can be derived analytically from the orbital parameters and the grid spacing, without computing the function on all $2^n$ grid points. This yields fully $O(n)$ state preparation: $O(n)$ classical operations to determine the tensor elements, followed by $O(n)$ quantum gates to execute the circuit.

\subsubsection{Exponential functions}

The simplest case is a single-sided exponential $f(x) = e^{-\zeta x}$ on a grid $x = \sum_k s_k \cdot 2^{n-1-k} \cdot \Delta x$ with $s_k \in \{0,1\}$. The exponential of a sum  factorizes as a product over bits:
\begin{equation}
e^{-\zeta x} = \prod_k e^{-\zeta \cdot s_k \cdot 2^{n-1-k} \cdot \Delta x}.
\end{equation}
Each factor depends on a single bit $s_k$, giving $\chi = 1$ (product state). The circuit consists of $n$ independent single-qubit $R_y$ rotations with angles determined by $\zeta$ and $\Delta x$.

\subsubsection{Piecewise exponentials: 1s STO}

The 1s orbital in one dimension, $f(x) = e^{-\zeta|x-R|}$ consists of a rising exponential for $x < R$ and a decaying exponential for $x > R$. A single qubit in superposition controls which branch is active, giving $\chi = 2$. The circuit is constructed by hand: controlled rotations separate the two branches, each of which is a product state. No MPS decomposition or SVD is required.

The derivative $d/dx[e^{-\zeta|x-R|}] = \pm\zeta \cdot e^{-\zeta|x-R|}$ has the same structure with a sign reversal  at the center; the derivative state is obtained by adding a single $Z$ gate  to the 1s preparation circuit.

\subsubsection{Linear \texorpdfstring{$\times$}{x} exponential: 2s STO}

The one-dimensional 2s STO $f(x) = x \cdot e^{-\zeta x}$ has $\chi = 2$. Since $x = \sum_k s_k \cdot 2^{n-1-k}$ is a sum over bits while $e^{-\zeta x} = \prod_k e^{-\zeta \cdot s_k \cdot 2^{n-1-k}}$ is a product, their product has MPS transfer matrices:
\begin{equation}
A^{[k]0} = I_2, \quad A^{[k]1} = e_k \begin{pmatrix} 1 & w_k \\ 0 & 1 \end{pmatrix},
\end{equation}
where $e_k = e^{-\zeta \cdot 2^{n-1-k} \cdot \Delta x}$ and $w_k = 2^{n-1-k}/\mathcal{N}$. The normalization $\mathcal{N}$ is computed in $O(n)$ by contracting $\chi^2 \times \chi^2 = 4 \times 4$ transfer matrices.

\subsubsection{Hydrogen 2s orbital}

The one-dimensional radial part $f(r) = (2 - r/a) \cdot e^{-r/2a}$ decomposes as the difference of a $\chi = 1$ term  (constant $\times$ exponential) and a $\chi = 2$ term  (linear $\times$ exponential). The direct sum gives $\chi = 3$, with $3 \times 3$  upper-triangular transfer matrices:
\begin{equation}
A^{[k]0} = I_3, \quad A^{[k]1} = e_k \begin{pmatrix} 1 & 0 & 0 \\ 0 & 1 & -p_k/a \\ 0 & 0 & 1 \end{pmatrix},
\end{equation}
where $e_k = e^{-2^{n-1-k}/(2a)}$ and $p_k = 2^{n-1-k}$. The left boundary vector $(2, 1, 0)$ and right boundary vector $(1, 0, 1)^T$ recover $f(r)$ through the matrix product. The node at $r = 2a$ is captured by the algebraic structure without special treatment.

\begin{table*}[t]
\caption{\label{tab:1d_mps} Analytical MPS constructions for 1D orbital functions.
Bond dimension $\chi = d+1$ for polynomial prefactor of degree $d$;
normalization computed via $O(n)$ transfer-matrix contraction.}
\begin{ruledtabular}
\begin{tabular}{lclcc}
Function & $\chi$ & Transfer matrix & Analytical & Norm \\
\hline
$e^{-\zeta|x-R|}$ (1s) & 2 & $2\!\times\!2$ diagonal & Yes & $O(n)$ \\
$\pm\zeta e^{-\zeta|x-R|}$ ($\partial_x$1s) & 2 & $2\!\times\!2$ $+$ $Z$ gate & Yes & $O(n)$ \\
$xe^{-\zeta x}$ (2s STO) & 2 & $2\!\times\!2$ upper tri. & Yes & $O(n)$ \\
$(2-x/a)e^{-x/2a}$ (H\,2s) & 3 & $3\!\times\!3$ ($\chi{=}1{\oplus}2$) & Yes & $O(n)$ \\
$x(2-x/a)e^{-x/2a}$ (2s+Jac.) & 3 & $3\!\times\!3$ upper tri. & Yes & $O(n)$ \\
$V(x)\psi(x)$ ($V\!\cdot\!\psi$) & 11 & numerical SVD & No & $O(2^n)$ \\
\end{tabular}
\end{ruledtabular}
\end{table*}

\subsubsection{Degree-2 polynomial \texorpdfstring{$\times$}{x} exponential}

The function $f(r) = r(2 - r/a) \cdot e^{-r/2a} = (2r - r^2/a) \cdot e^{-r/2a}$ is a degree-2 polynomial $\times$ exponential with $\chi = 3$. The physical context in which this function arises is discussed in Sec.~\ref{sec:3d_prep}. The three bond states track the running polynomial: state~0 carries the constant (1), state~1 the linear sum ($x$), and state~2 the quadratic sum ($x^2$). The transfer matrices are:
\begin{equation}
A^{[k]0} = I_3, \quad A^{[k]1} = e_k \begin{pmatrix} 1 & p_k & p_k^2 \\ 0 & 1 & 2p_k \\ 0 & 0 & 1 \end{pmatrix},
\end{equation}
with left boundary $(1, 0, 0)/\mathcal{N}$ and right boundary $(0, 2, -1/a)^T$. The normalization is computed in $O(n)$ via $9 \times 9$ transfer matrix contraction.

\subsubsection{General pattern}

For a one-dimensional function of the form $p_d(x) \cdot e^{-\zeta x}$ where $p_d$ is a polynomial of degree $d$, the MPS bond dimension is $\chi = d + 1$. The transfer matrices are $(d{+}1) \times (d{+}1)$ upper triangular, with entries determined by the binomial expansion of $(\sum s_k \cdot 2^{n-1-k})^m$ for $m = 0, \ldots, d$. All tensors, normalization constants, and circuit parameters are computable in $O(n)$ classical operations with $O(1)$-size matrices.

\subsection{State preparation in three dimensions}
\label{sec:3d_prep}

\subsubsection{Spherical coordinates}

For s-orbitals, the 3D wavefunction $\psi(r,\theta,\varphi) = R(r) \cdot Y_0^0$ is spherically symmetric. The angular part is a constant and requires no quantum encoding. The state is prepared on a single radial register using the 1D analytical MPS constructions of Sec.~\ref{sec:analytical_mps}.

In spherical coordinates, overlap integrals include the volume element $r^2 dr$. For s-orbitals, one factor of $r$ is absorbed into the radial wavefunction $R(r)$, leaving a Jacobian factor of $r$ in the integrand. Computing overlap integrals therefore requires preparing states that include this factor---for example, the 2s radial integrand $r(2 - r/a) \cdot e^{-r/2a}$, whose analytical MPS construction is given in Sec.~\ref{sec:analytical_mps} ($\chi = 3$, degree-2 polynomial $\times$ exponential).

The preparation cost is identical to the 1D case: $O(n)$ gates with constant bond dimension. Orthogonality between orbitals of different principal quantum number (e.g.\ $\braket{1s|2s} = 0$) is enforced by the radial wavefunctions and can be verified using the same compute/uncompute overlap circuit.

\subsubsection{Cartesian coordinates}

For multi-center problems, Cartesian coordinates provide a natural grid on which orbitals centered at different positions can be represented without coordinate transformations. The 3D function $e^{-\zeta\sqrt{x^2+y^2+z^2}}$ is encoded on three registers $\ket{x}\ket{y}\ket{z}$ as a single MPS over all $3n$ qubits via numerical right-canonical SVD (Sec.~\ref{sec:mps}). No analytical MPS construction exists because the coupling through $r = \sqrt{x^2+y^2+z^2}$ prevents factorization across coordinates.

The qubit ordering within the MPS chain is $\ket{x}\ket{y}\ket{z}$ (grouped by coordinate), with $x$ bits as the most significant and $z$ bits as the least significant. This ordering was found to produce substantially lower bond dimensions than the interleaved alternative $\ket{x_1 y_1 z_1 x_2 y_2 z_2 \ldots x_n y_n z_n}$ (Sec.~\ref{sec:entanglement}). The bond dimension and its scaling behavior under grid resolution, orbital decay constant, and SVD truncation threshold are analyzed in detail in Sec.~\ref{sec:entanglement}.

\subsection{One-electron integrals}
\label{sec:integrals}

\subsubsection{Overlap integral}

The overlap $S_{AB} = \braket{\psi_A|\psi_B}$ between two amplitude-encoded orbitals is evaluated using the compute/uncompute method on a single register:
\begin{equation}
\braket{\psi_A|\psi_B} = \bra{0} U_A^\dagger U_B \ket{0}.
\end{equation}
The probability of measuring $\ket{0\ldots0}$ gives $P(0) = |\braket{\psi_A|\psi_B}|^2$. The physical overlap is:
\begin{equation}
S_{AB} = \mathcal{N}_A \cdot \mathcal{N}_B \cdot (\pm\sqrt{P(0)}) \cdot \Delta x,
\end{equation}
where $\mathcal{N}_A$, $\mathcal{N}_B$ are the discrete L2 norms and the sign is determined from the physics of the orbital pair (positive for nodeless orbitals, requiring a Hadamard test for orbitals with nodes).

For overlap integrals, the normalization factors simplify. Since both $\psi_A$ and $\psi_B$ are normalized to the same physical convention, $\mathcal{N}_A \cdot \mathcal{N}_B \cdot \Delta x$ appears identically in both the numerator and the denominator of the normalized overlap $S_{AB} = \int \psi_A^* \psi_B\, dx \,/\, (\sqrt{\int|\psi_A|^2\, dx} \cdot \sqrt{\int|\psi_B|^2\, dx})$. All prefactors cancel, and the physical overlap reduces to $S_{AB} = \pm\sqrt{P(0)}$. The full normalization chain~\eqref{eq:physical_integral} is required only for matrix elements of non-unitary operators (Sec.~\ref{sec:integrals}, Nuclear attraction integral), where the bra and ket states have different norms.

This method uses $n$ physical qubits plus $\lceil\log_2(\chi_{\max})\rceil$ bond qubits, where $\chi_{\max}$ is the larger bond dimension of the two orbitals. The circuit depth is $2n$ unitary gates ($n$ for $U_B$, $n$ for $U_A^\dagger$), with possible gate cancellations at the junction.

\subsubsection{Kinetic energy integral}

The kinetic energy matrix element $T_{AB} = -\tfrac{1}{2}\bra{\psi_A}\tfrac{d^2}{dx^2}\ket{\psi_B}$ is computed using the identity $T_{AB} = \tfrac{1}{2}\braket{\partial\psi_A/\partial x|\partial\psi_B/\partial x}$, obtained by integration by parts with vanishing boundary terms. This reduces the kinetic energy to an overlap of derivative states.

For the 1s STO, the derivative is $d/dx[e^{-\zeta|x-R|}] = \pm\zeta \cdot e^{-\zeta|x-R|}$, differing from the original function only by a sign reversal at the center. The derivative state is prepared by the same circuit as the 1s orbital with one additional $Z$ gate per preparation. The kinetic energy circuit therefore has the same two-qubit gate count as the overlap circuit, with only two additional single-qubit gates.

\subsubsection{Nuclear attraction integral}

The nuclear attraction matrix element $V_{AB} = \bra{\psi_A}(-Z/|x - R_C|)\ket{\psi_B}$ cannot be evaluated by inserting a unitary operator between compute and uncompute steps, because the potential $V(x) = -Z/|x - R_C|$ is a diagonal but non-unitary operator---it multiplies each basis state by a different real number, changing the norm.

Instead, we encode the product function $\varphi_B(x) = V(x) \cdot \psi_B(x)$ as a new amplitude-encoded state and reduce the matrix element to an overlap:
\begin{equation}
V_{AB} = \mathcal{N}_A \cdot \mathcal{N}_\varphi \cdot \braket{\psi_A^{\text{norm}}|\varphi_B^{\text{norm}}} \cdot \Delta x,
\end{equation}
where $\mathcal{N}_\varphi = \sqrt{\sum_j |V(x_j) \cdot \psi_B(x_j)|^2}$ is the norm of the product function. The MPS decomposition of $\varphi_B$ is obtained by numerical SVD; its bond dimension saturates at $\chi = 11$ for the 1D Coulomb potential (Sec.~\ref{sec:entanglement}), confirming that the encoding is efficient.

For functions whose MPS cannot be derived analytically, TT-cross interpolation~\cite{Oseledets2010} provides a scalable alternative to full-grid evaluation: the MPS tensors are constructed from $O(n \cdot \chi^2)$ adaptively chosen function evaluations rather than all $2^n$ grid points.

\subsection{Experimental setup}
\label{sec:setup}

Hardware experiments were performed on IBM Heron-generation quantum processors \cite{IBMQuantum}: ibm\_kingston and ibm\_marrakesh, both featuring 156~qubits with CZ as the native two-qubit gate. Circuits were transpiled using Qiskit \cite{Qiskit} at optimization level~3 with multiple transpilation attempts to select the lowest gate count.

Two measurement primitives were used. The Sampler primitive executes the circuit and returns the probability distribution over measurement outcomes, from which $P(0) = |\braket{0\ldots0|\psi}|^2$ is extracted directly. The Estimator primitive evaluates the expectation value of an observable $O = \ket{0\ldots0}\bra{0\ldots0} \otimes I_{\text{anc}}$, constructed as a \texttt{SparsePauliOp}, which gives $P(0)$ with access to error mitigation techniques not available to the Sampler.

Several error mitigation strategies were characterized: Pauli twirling, measurement error mitigation, dynamical decoupling, and Zero-Noise Extrapolation (ZNE)~\cite{Temme2017,Li2017}. ZNE is available only through the Estimator primitive; it estimates the zero-noise expectation value by running the circuit at progressively amplified noise levels (factors $[1, 2, 3]$) and extrapolating to the zero-noise limit with a linear fit. Of the techniques tested, the Estimator with ZNE consistently produced the best results and is used for all hardware values reported in this work.

Noisy simulations were performed using the \texttt{fake\_sherbrooke} noise model in Qiskit Aer, which emulates the noise characteristics of an IBM Eagle-generation processor. It uses ECR as the native two-qubit gate. The noise model includes depolarizing errors, thermal relaxation, and readout errors calibrated to the physical device.

%% ============================================================
%% SECTION III: RESULTS
%% ============================================================

\section{Results}
\label{sec:results}

\subsection{One-dimensional integral evaluation}
\label{sec:1d_results}

We evaluate overlap, kinetic energy, and nuclear attraction integrals between 1s Slater-type orbitals in one dimension at the equilibrium bond distance of H$_2$ ($d = 1.4$~a.u.). The simulated space spans 16~a.u.\ discretized on a grid of $2^n$ points, with $n$ ranging from 4 to 10 qubits. Since the discretized grid generally does not contain a point at exactly 1.4~a.u., we take the nearest grid distance for each qubit count and compute the analytical reference at that discretized distance.

\subsubsection{Overlap integral}

\begin{figure}
\includegraphics[width=\columnwidth]{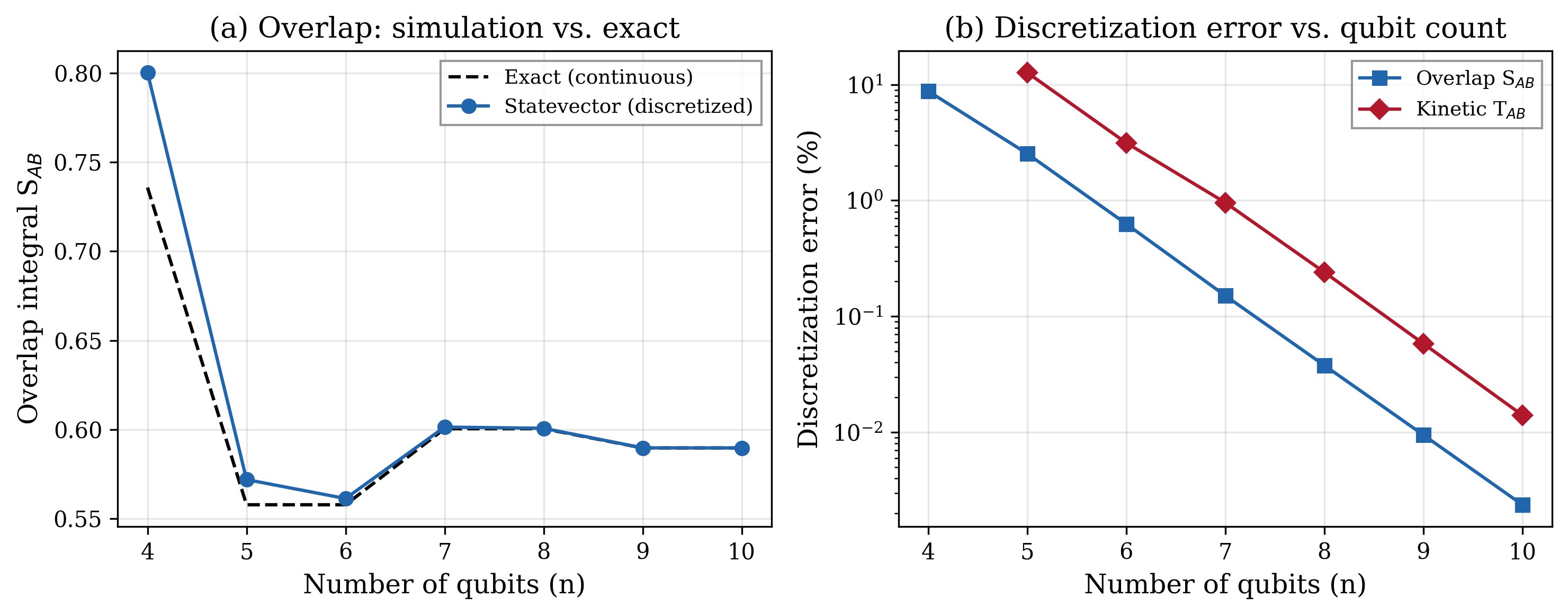}
\caption{\label{fig:convergence} Discretization convergence of 1D integrals. (a)~Overlap integral converging to the exact value $S(1.4) = 0.592$ (dotted line). (b)~Discretization error versus qubit count for overlap and kinetic energy integrals.}
\end{figure}

The overlap integral $S_{AB} = \braket{\psi_A|\psi_B}$ is computed via the compute/uncompute circuit: the state $\ket{0}$ is evolved to $\ket{\psi_B}$ by the preparation unitary $U_B$, then $U_A^\dagger$ is applied, and the probability of measuring $\ket{0\ldots0}$ yields $P(0) = |S_{AB}|^2$. For 1s orbitals, all amplitudes are positive, so $S_{AB} = \sqrt{P(0)}$ without sign ambiguity.

Figure~\ref{fig:convergence}(a) shows the statevector simulation result converging to the exact continuous integral $S(1.4) = 0.592$ as the qubit count increases. At 4~qubits per register, the discretization error is 8.8\%, dominated by the coarse grid mapping 1.4~a.u.\ to 1.0~a.u. By 8~qubits the discretization error drops to 0.04\%, and at 10~qubits it is below 0.01\% [Fig.~\ref{fig:convergence}(b)]. The statevector result agrees with the exact value at the discretized distance to machine precision at all qubit counts, confirming the correctness of the MPS-based state preparation.

\subsubsection{Kinetic energy integral}

The kinetic energy matrix element $T_{AB} = -\tfrac{1}{2}\bra{\psi_A}\tfrac{d^2}{dx^2}\ket{\psi_B}$ is evaluated using the derivative encoding of the 1s orbital. Integration by parts gives the equivalent form $T_{AB} = \tfrac{1}{2}\braket{\partial\psi_A/\partial x|\partial\psi_B/\partial x}$, which reduces the computation to an overlap of derivative states. Since $d/dx[e^{-\zeta|x-R|}] = \pm\zeta \cdot e^{-\zeta|x-R|}$ with a sign reversal at the center, the derivative state is obtained from the 1s preparation circuit by adding a single $Z$ gate that flips the sign of the decaying branch. The kinetic energy circuit is therefore identical to the overlap circuit except for one additional single-qubit gate per preparation.

The discretization error for kinetic energy is consistently larger than for the overlap at the same qubit count: 12.7\% at 5~qubits versus 2.5\% for the overlap [Fig.~\ref{fig:convergence}(b)]. This is expected because the derivative amplifies grid-resolution effects---the slope of the exponential changes more rapidly than the function itself. At 8~qubits the discretization error drops to 0.24\%, confirming that the circuit is correct and the error is purely a grid resolution effect.

\subsubsection{Nuclear attraction integral}

The nuclear attraction matrix element $V_{AB} = \bra{\psi_A}(-Z/|x-R_C|)\ket{\psi_B}$ is computed by encoding the product function $\varphi_B(x) = V(x) \cdot \psi_B(x)$ as a separate MPS state and reducing the matrix element to an overlap: $V_{AB} = \mathcal{N}_A \cdot \mathcal{N}_\varphi \cdot \braket{\psi_A^{\text{norm}}|\varphi_B^{\text{norm}}} \cdot \Delta x$, where $\mathcal{N}_A$ and $\mathcal{N}_\varphi$ are the discrete L2 norms computed classically during MPS construction and $\Delta x$ is the grid spacing.

The bond dimension of $\varphi_B(x) = -Z \cdot e^{-\zeta|x-R_B|}/|x-R_C|$ saturates at $\chi = 11$ independent of qubit count (Table~\ref{tab:1d_mps}), confirming that the MPS encoding of the potential-weighted orbital is efficient. Unlike the pure orbital ($\chi = 2$), no analytical MPS construction exists for $V \cdot \psi$---the tensors are obtained by numerical SVD. However, the saturation of $\chi$ means that TT-cross interpolation~\cite{Oseledets2010} can construct the MPS in $O(n \cdot \chi^2)$ function evaluations, avoiding the exponential cost of full grid sampling.

The 1D Coulomb potential $1/|x - R_C|$ is not integrable: the integral $\int e^{-2\zeta|x-R|}/|x - R_C|\, dx$ diverges logarithmically for all orbital configurations, including the off-center case. The discretized matrix element computed on a finite grid is well-defined but grid-dependent---it does not converge to a finite value with increasing resolution. We present the nuclear attraction computation as a methodological demonstration: the $V \cdot \psi$ encoding, MPS decomposition, and overlap extraction pipeline produces exact agreement with the corresponding discretized sum at all tested qubit counts (4--8), with circuit fidelity of 1.0 in statevector simulation. This validates the approach for application in three dimensions, where the volume element $r^2 dr$ renders the Coulomb integral finite. The 3D nuclear attraction integral is left for future work.

The transpiled circuit for the nuclear attraction computation contains 8,486~ECR gates (Table~\ref{tab:circuits}), arising from the $\chi = 11$ bond dimension of the $V \cdot \psi$ state. This is infeasible on current hardware.

\begin{table*}[t]
\caption{\label{tab:circuits} Transpiled circuit metrics.
1D circuits transpiled for ibm\_kingston / ibm\_marrakesh (CZ native gate);
3D circuits transpiled for fake\_sherbrooke noise model (ECR native gate).
Values are representative of the best result over multiple transpilation attempts.
The interleaved qubit ordering produces $\sim\!4\times$ more gates than the
grouped $\ket{x}\ket{y}\ket{z}$ ordering.}
\begin{ruledtabular}
\begin{tabular}{lrrrl}
Circuit & Qubits & 2-qubit gates & Depth & Backend \\
\hline
\multicolumn{5}{l}{\textit{1D circuits}} \\
Overlap, $n=5$ & 5 & $\sim\!105$ CZ & $\sim\!266$ & ibm\_kingston \\
Overlap, $n=6$ & 6 & $\sim\!134$ CZ & $\sim\!305$ & ibm\_kingston \\
Overlap, $n=8$ & 8 & $\sim\!250$ CZ & $\sim\!550$ & ibm\_kingston \\
Kinetic, $n=5$ & 5 & $\sim\!107$ CZ & $\sim\!264$ & ibm\_kingston \\
$\braket{1s|2s}$ spherical, $4$q/coord & 4 & 149 ECR & 669 & fake\_sherbrooke \\
$\braket{1s|2s}$ spherical, $8$q/coord & 8 & 310 ECR & 1381 & fake\_sherbrooke \\
$V\!\cdot\!\psi$, $n=6$ & 9 & 8,486 ECR & 36,624 & (sim only) \\[3pt]
\multicolumn{5}{l}{\textit{3D Cartesian circuits (threshold $10^{-12}$)}} \\
1s prep, $3$q/coord & 9 & 9,552 ECR & 40,541 & (sim only) \\
1s overlap, $3$q/coord & 9 & 18,974 ECR & 80,598 & (sim only) \\
2s prep, $4$q/coord & 12 & 221,063 ECR & 927,132 & (sim only) \\
1s overlap, interleaved, $3$q/coord & 9 & 39,026 ECR & 171,454 & (sim only) \\
\end{tabular}
\end{ruledtabular}
\end{table*}

\subsection{Three-dimensional overlap integrals}
\label{sec:3d_results}

To demonstrate the extension of the methodology to three dimensions, we encode 1s and 2s hydrogen orbitals and compute overlap integrals via statevector simulation.

\subsubsection{Spherical coordinates: same-center orthogonality}

In spherical coordinates, the radial parts of 1s and 2s orbitals are encoded on a single register using the analytical MPS constructions described in Sec.~\ref{sec:analytical_mps}. Since the angular parts are identical for s-orbitals, the orthogonality $\braket{1s|2s}$ is determined entirely by the radial overlap. We simulate this on a 32~a.u.\ radial grid at 4 to 8~qubits.

The statevector result converges rapidly toward zero: $\braket{1s|2s} = 0.183$ at 4~qubits, 0.019 at 5~qubits, 0.001 at 6~qubits, and $6 \times 10^{-6}$ at 8~qubits (Table~\ref{tab:3d_overlap}). The residual at low qubit counts reflects discretization error in the radial grid, not a failure of orthogonality. The transpiled circuits are modest---149~ECR gates at 4~qubits, 310~ECR gates at 8~qubits---confirming the efficiency of separable coordinate encodings.

\begin{table*}[t]
\caption{\label{tab:3d_overlap} Three-dimensional overlap integrals.
\textit{Spherical coordinates}: radial overlap $\braket{1s|2s}$ (same-center; exact
value is zero). \textit{Cartesian coordinates}: multi-center overlaps at $d = 1.4$~a.u.\
(H$_2$ equilibrium). All values are amplitudes ($\sqrt{P(0)}$), not probabilities.
Error $= |\text{statevec} - \text{exact}|/|\text{exact}|$ for non-zero exact values;
deviation from zero otherwise.}
\begin{ruledtabular}
\begin{tabular}{lcrrrrr}
Integral & Space (a.u.) & $n$/coord & Total qubits & Exact & Statevec. & Error \\
\hline
\multicolumn{7}{l}{\textit{Spherical coordinates, $\braket{1s|2s}$ same-center}} \\
$\braket{1s|2s}$ & 32 & 4 & 4 & 0.0000 & 0.1826 & (0.183) \\
 & 32 & 5 & 5 & 0.0000 & 0.0187 & (0.019) \\
 & 32 & 6 & 6 & 0.0000 & 0.00138 & (0.0014) \\
 & 32 & 8 & 8 & 0.0000 & $6.0\!\times\!10^{-6}$ & ($6\!\times\!10^{-6}$) \\[3pt]
\multicolumn{7}{l}{\textit{Cartesian coordinates, $d = 1.4$~a.u.\ (threshold $10^{-12}$)}} \\
$\braket{1s_A|1s_B}$ & 16 & 3 & 9  & 0.7529 & 0.5704 & 24.24\% \\
 & 16 & 4 & 12 & 0.7529 & 0.7211 &  4.23\% \\
 & 16 & 5 & 15 & 0.7529 & 0.7493 &  0.48\% \\
 & 16 & 6 & 18 & 0.7529 & 0.7528 &  0.02\% \\[3pt]
$\braket{2s_A|2s_B}$ & 16 & 4 & 12 & 0.9333 & 0.9231 & 1.09\% \\
 & 16 & 5 & 15 & 0.9333 & 0.9305 & 0.29\% \\
 & 16 & 6 & 18 & 0.9333 & 0.9313 & 0.21\% \\
 & 16 & 7 & 21 & 0.9333 & 0.9313 & 0.21\% \\[3pt]
$\braket{1s_A|2s_B}$ (thr.\ $10^{-6}$) & 32 & 6 & 18 & $-0.1030$ & $-0.1022$ & 0.78\% \\
 & 64 & 7 & 21 & $-0.1030$ & $-0.1021$ & 0.79\% \\
\end{tabular}
\end{ruledtabular}
\end{table*}

\subsubsection{Cartesian coordinates: multi-center overlaps}

\begin{figure}
\includegraphics[width=\columnwidth]{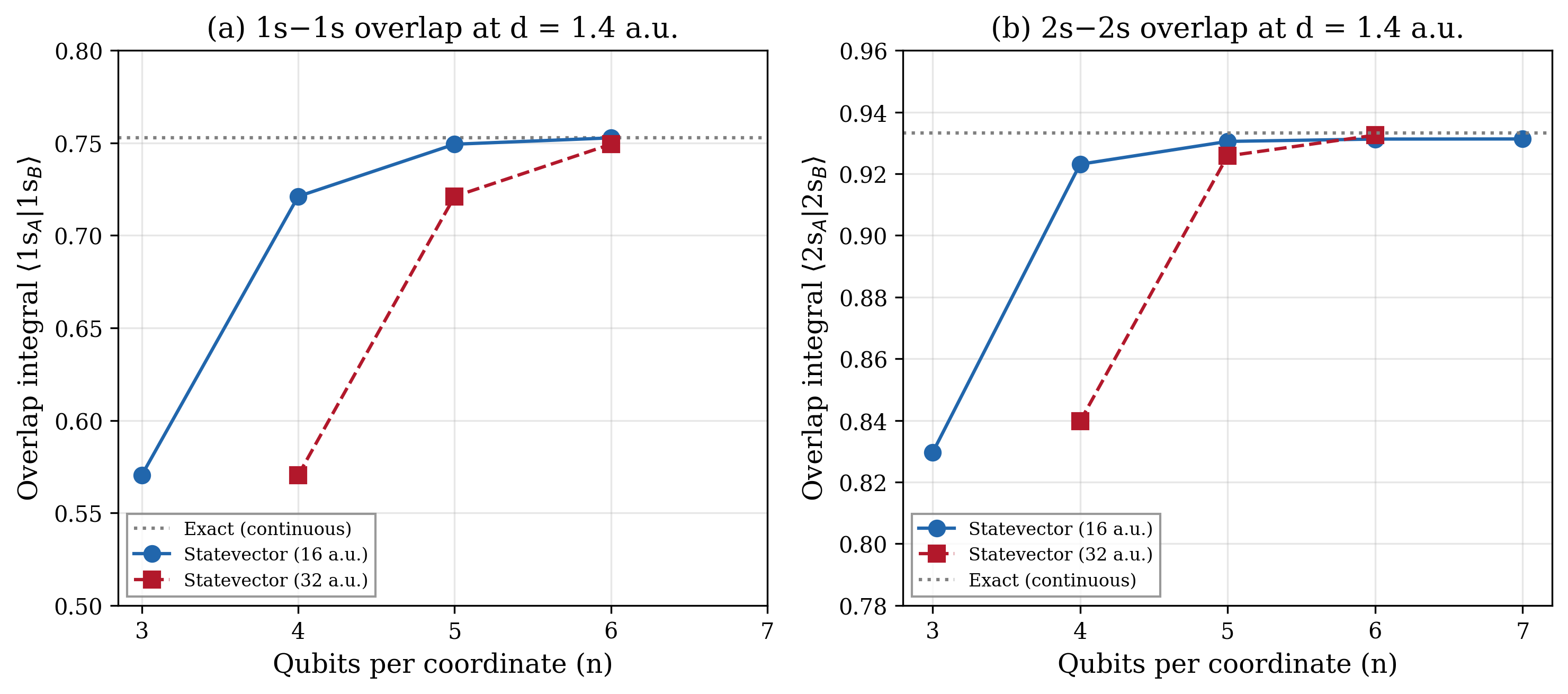}
\caption{\label{fig:3d_overlap} 3D overlap convergence. (a)~$\braket{1s_A|1s_B}$ at $d=1.4$~a.u. (b)~$\braket{2s_A|2s_B}$ at $d=1.4$~a.u. Dotted line: exact continuous value.}
\end{figure}

For the Cartesian representation, 1s and 2s orbitals are encoded on three registers $\ket{x}\ket{y}\ket{z}$ using MPS decomposition over all $3n$ qubits (Sec.~\ref{sec:mps}). The simulated space spans 16 to 64~a.u.\ per coordinate, with 3 to 7~qubits per coordinate.

Table~\ref{tab:3d_overlap} summarizes the results. For the 1s--1s overlap at the H$_2$ bonding distance (1.4~a.u.), the discretization error decreases from 24\% at 3~qubits per coordinate ($8^3 = 512$ grid points) to 0.02\% at 6~qubits per coordinate ($64^3 \approx 2.6 \times 10^5$ grid points). The 2s--2s overlap converges more rapidly, reaching 0.21\% error at 6~qubits per coordinate; this faster convergence reflects the slower spatial decay of the 2s orbital, which is better resolved on a given grid. The 2s integrals require a larger simulated space (64~a.u.) for convergence, consistent with the more extended spatial support of the 2s orbital.

The cross-orbital overlap $\braket{1s_A|2s_B}$ at 1.4~a.u.\ separation yields $-0.1022$ versus the exact value $-0.1030$ (0.78\% error) at 6~qubits per coordinate on a 32~a.u.\ grid. The same-center orthogonality $\braket{1s|2s}$ at zero separation gives a residual of 0.004 in Cartesian coordinates at 6~qubits per coordinate, consistent with the discretization error observed in the spherical encoding at comparable resolution.

These results confirm that the MPS-based amplitude encoding produces correct multi-center overlap integrals in three dimensions. However, the transpiled circuit depth is substantial: even at 3~qubits per coordinate (9 total qubits), the 1s overlap circuit contains 18,974~ECR gates (Table~\ref{tab:circuits}). At 4~qubits per coordinate the 2s preparation alone requires 221,063~ECR gates. These circuits are validated in simulation but lie beyond the reach of current noisy intermediate-scale hardware.

Figure~\ref{fig:3d_overlap} shows the convergence of the 3D overlap integrals with increasing qubit count. The convergence behavior at 32~a.u.\ space range is identical to that at 16~a.u.\ but shifted by one qubit---consistent with the fact that doubling the simulated space at fixed qubit count halves the grid resolution.

\subsection{Entanglement analysis}
\label{sec:entanglement}

\subsubsection{Bond dimensions of 1D orbital encodings}

Table~\ref{tab:1d_mps} summarizes the MPS bond dimensions for all 1D functions studied. The 1s orbital $e^{-\zeta|x-R|}$ has $\chi = 2$, with the bond state distinguishing the rising and decaying branches on either side of the center. For polynomial $\times$ exponential functions of the form $p(x) \cdot e^{-\zeta x}$, the bond dimension equals $d + 1$ where $d$ is the polynomial degree: $x \cdot e^{-\zeta x}$ has $\chi = 2$, and $x(2 - x/a) \cdot e^{-x/2a}$ has $\chi = 3$. In all cases the MPS tensors are derivable analytically from the orbital parameters and the grid spacing, requiring $O(n)$ classical operations and no sampling of the $2^n$-point grid (Sec.~\ref{sec:analytical_mps}).

The nuclear attraction product $V \cdot \psi$ breaks this pattern: its bond dimension saturates at $\chi = 11$, requiring numerical SVD for tensor construction. The saturation reflects the fact that the Coulomb singularity, while not analytically factorizable, has bounded complexity on the binary grid---the singularity affects only $O(1)$ grid points near $R_C$, and the surrounding smooth exponential decay keeps the effective rank finite.

\subsubsection{Bond dimensions of 3D Cartesian encodings}

\begin{figure*}
\includegraphics[width=\textwidth]{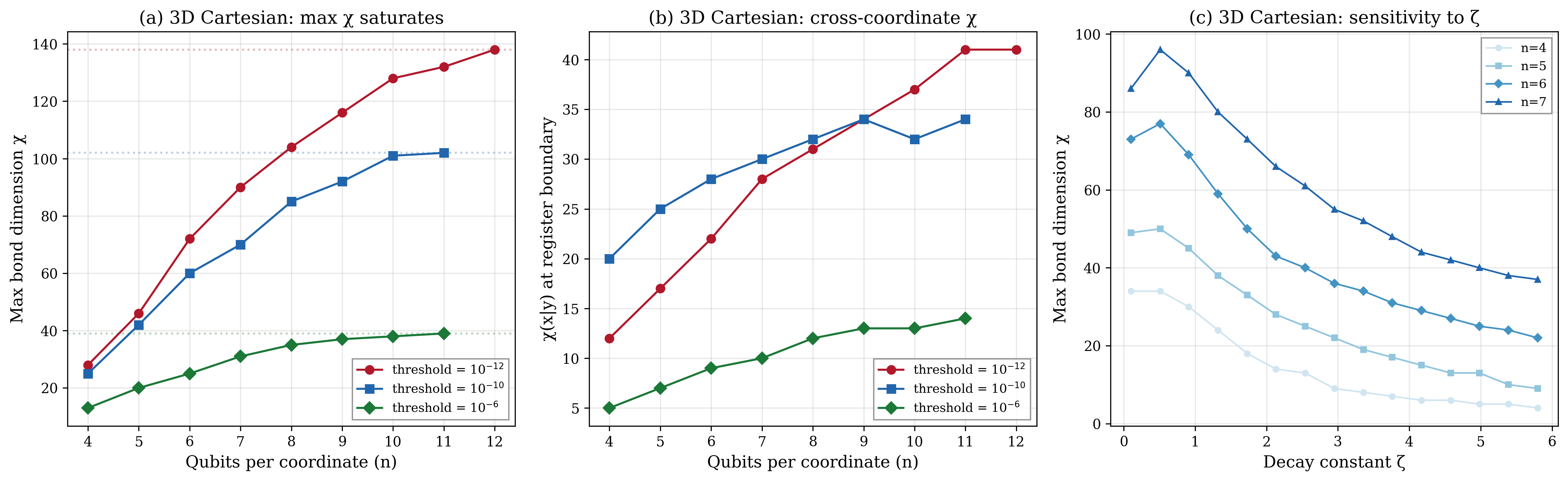}
\caption{\label{fig:bond_dim} Bond dimension scaling of 3D Cartesian encodings. (a)~3D Cartesian $\chi_{\max}$ versus qubits per coordinate at three SVD thresholds, showing saturation. (b)~Cross-coordinate $\chi(x|y)$, also saturating. (c)~Sensitivity of $\chi_{\max}$ to decay constant $\zeta$.}
\end{figure*}

The entanglement structure of three-dimensional orbitals encoded on a Cartesian grid $\ket{x}\ket{y}\ket{z}$ is characterized by two quantities: the maximum bond dimension $\chi_{\max}$ along the MPS chain, and the cross-coordinate Schmidt rank $\chi(x|y)$ at the register boundary. We study the scaling of both quantities under three variables: grid resolution (qubits per coordinate), orbital decay constant $\zeta$, and SVD truncation threshold.

\emph{Grid resolution} [Fig.~\ref{fig:bond_dim}(a,b); Table~\ref{tab:3d_bonds}]. Both $\chi_{\max}$ and $\chi(x|y)$ grow with qubit count but saturate. For the 1s orbital at $\zeta = 1$ on a 32~a.u.\ grid, $\chi_{\max}$ increases from 28 ($n=4$) to 138 ($n=12$), with successive increments $\Delta\chi$ peaking at $n = 6$ (+26) and then decreasing monotonically to +6 at $n = 12$. The cross-coordinate entanglement $\chi(x|y)$ grows from 12 to 41 over the same range, and stops entirely between $n = 11$ and $n = 12$: $\chi(x|y) = 41$ at both qubit counts. The cross-coordinate entanglement saturates before the intra-coordinate structure, indicating that the coupling between coordinate registers through $r = \sqrt{x^2 + y^2 + z^2}$ is fully resolved at moderate grid density, while the fine-scale structure within each register continues to develop. Self-consistency is confirmed by the observation that the 64~a.u.\ data at $n$ qubits matches the 32~a.u.\ data at $n - 1$ qubits across all qubit counts---doubling the space at fixed qubit count is equivalent to halving the resolution.

\emph{Decay constant} [Fig.~\ref{fig:bond_dim}(c)]. Larger $\zeta$ reduces both $\chi_{\max}$ and $\chi(x|y)$ at all grid sizes. At $n = 7$ qubits per coordinate, $\chi_{\max} = 87$ for $\zeta = 1$ versus 36 for $\zeta = 6$. Faster decay localizes the orbital on fewer grid points, reducing the effective entanglement. For heavier atoms, where orbital exponents are larger, the Cartesian encoding cost will be correspondingly lower.

\emph{Truncation threshold} [Fig.~\ref{fig:bond_dim}(a,b); Table~\ref{tab:3d_bonds}]. Reducing the SVD threshold from $10^{-12}$ to $10^{-6}$ cuts $\chi_{\max}$ from 132 to 39 at $n = 11$---a factor of 3.4---while reducing the number of required bond qubits from 8 to 6. The impact on integral accuracy is negligible: the 3D overlap integrals computed with threshold $10^{-6}$ agree with those at $10^{-12}$ to within the discretization error (Table~\ref{tab:3d_overlap}). This provides a practical tuning parameter that trades precision of the amplitude encoding against circuit resources.

\begin{table*}[t]
\caption{\label{tab:3d_bonds} Bond dimension scaling for 3D Cartesian MPS encoding
(hydrogen 1s orbital, $\zeta=1$, 32~a.u.\ grid).
$\chi_{\max}$: maximum bond dimension along the $3n$-qubit MPS chain (bottleneck
occurs in the $y$ register). $\chi(x|y)$: Schmidt rank at the $x$--$y$ register
boundary. $\Delta$: increment from the previous row.
Both quantities saturate with increasing $n$.}
\begin{ruledtabular}
\begin{tabular}{rrrrrrrr}
\multicolumn{2}{c}{} & \multicolumn{4}{c}{Threshold $10^{-12}$} & \multicolumn{2}{c}{Threshold $10^{-6}$} \\
\cline{3-6}\cline{7-8}
$n$/coord & $3n$ & $\chi_{\max}$ & $\Delta\chi_{\max}$ & $\chi(x|y)$ & $\Delta\chi(x|y)$ & $\chi_{\max}$ & $\chi(x|y)$ \\
\hline
 4 & 12 & 28  & ---  & 12 & --- & 13 & 5  \\
 5 & 15 & 46  & $+$18 & 17 & $+$5  & 20 & 7  \\
 6 & 18 & 72  & $+$26 & 22 & $+$5  & 25 & 9  \\
 7 & 21 & 90  & $+$18 & 28 & $+$6  & 31 & 10 \\
 8 & 24 & 104 & $+$14 & 31 & $+$3  & 35 & 12 \\
 9 & 27 & 116 & $+$12 & 34 & $+$3  & 37 & 13 \\
10 & 30 & 128 & $+$12 & 37 & $+$3  & 38 & 13 \\
11 & 33 & 132 & $+$4  & 41 & $+$4  & 39 & 14 \\
12 & 36 & 138 & $+$6  & 41 & $\phantom{+}$0  & --- & --- \\
\end{tabular}
\end{ruledtabular}
\end{table*}

\subsubsection{Bond profile structure}

\begin{figure*}
\includegraphics[width=\textwidth]{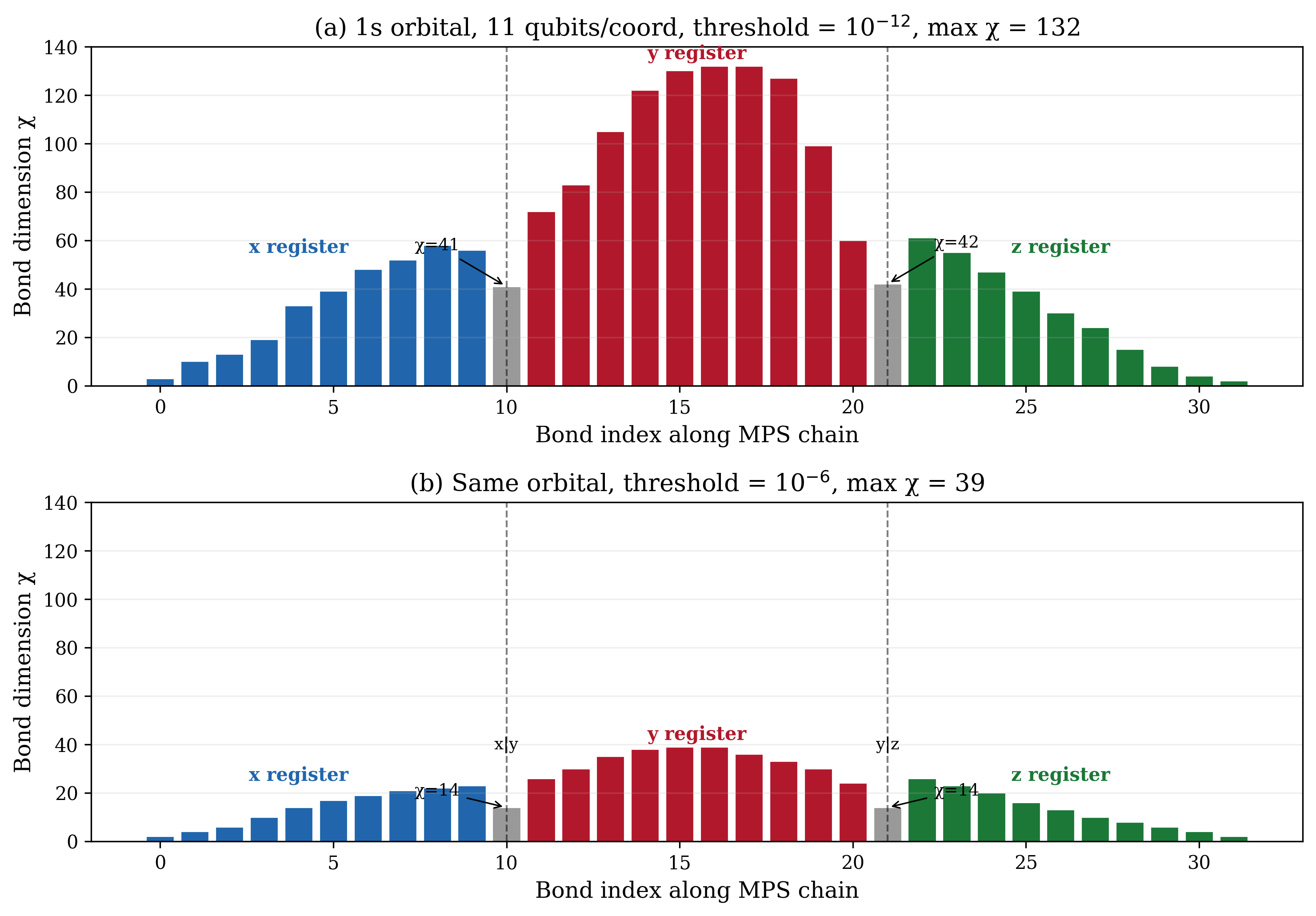}
\caption{\label{fig:bond_profile} Bond dimension profile across the $3n$-qubit MPS chain for the 1s orbital at 11~qubits per coordinate. (a)~Threshold $10^{-12}$: $y$-register bonds peak at $\chi = 132$ (bottleneck effect). (b)~Threshold $10^{-6}$: peak reduced to $\chi = 39$. Gray bars mark the $x|y$ and $y|z$ register boundaries.}
\end{figure*}

Figure~\ref{fig:bond_profile} shows the full bond dimension profile across the $3n$-qubit MPS chain for the 1s orbital at 11~qubits per coordinate. The profile reveals a characteristic arch shape: bonds within the $y$ register reach the maximum ($\chi = 132$ at threshold $10^{-12}$), while bonds within the $x$ and $z$ registers peak at approximately 58. This asymmetry is not physical---the function $e^{-\zeta r}$ is symmetric under $x \leftrightarrow z$---but structural. In the grouped qubit ordering $\ket{x}\ket{y}\ket{z}$, bonds within the $y$ register must carry all correlations between $x$ and $z$, acting as a bottleneck. The $x$ and $z$ bond profiles are mirror images of each other, confirming the underlying symmetry.

The cross-coordinate bonds ($\chi = 41$ at the $x|y$ and $y|z$ boundaries) are substantially smaller than the intra-register maximum, indicating that the dominant entanglement cost comes from resolving the radial structure $\sqrt{x^2 + y^2 + z^2}$ within each coordinate register, not from the coupling between coordinates.

\subsubsection{Qubit ordering}

We compared the grouped ordering $\ket{x}\ket{y}\ket{z}$ with the interleaved ordering $\ket{x_1 y_1 z_1 x_2 y_2 z_2 \ldots x_n y_n z_n}$. At 3~qubits per coordinate, the interleaved ordering produces a maximum bond dimension of 64 versus 28 for grouped, and the transpiled circuit contains 39,026~ECR gates versus 9,552---a factor of 4 increase (Table~\ref{tab:circuits}). At larger $n$ the interleaved $\chi_{\max}$ grows rapidly (332 at $n = 9$), confirming that the grouped ordering is the correct choice for this problem. The bottleneck effect in the $y$ register is the price paid for a globally more efficient MPS representation.

\subsubsection{Two-electron integrals}

The Coulomb interaction kernel $1/|x_1 - x_2|$ on two registers has full Schmidt rank (equal to the grid size $N$) across the $x_1|x_2$ bipartition. The full integrand $f_1(x_1) \cdot (1/|x_1 - x_2|) \cdot f_2(x_2)$, despite the exponential localization of the orbital charge distributions, retains near-full Schmidt rank across the $x_1|x_2$ bipartition at all grid sizes tested: $\chi(x_1|x_2)$ grows as approximately $0.85N$, where $N = 2^n$ is the grid size ($N = 8$ to 1024). The Fourier decomposition of $1/|x|$ requires $O(N)$ terms for convergence, with 99\% of the spectral weight distributed across 97\% of the modes. These findings indicate that efficient amplitude encoding of the two-electron integrand in one dimension is not feasible with the present approach. In three dimensions, the multipole expansion of $1/|\mathbf{r}_1 - \mathbf{r}_2|$ provides a separable decomposition into angular momentum channels, each of which factorizes into radial and angular functions amenable to the encoding methods developed here. This direction is left for future work.

\subsection{Hardware validation}
\label{sec:hardware}

\subsubsection{Overlap integral on hardware}

\begin{figure}
\includegraphics[width=\columnwidth]{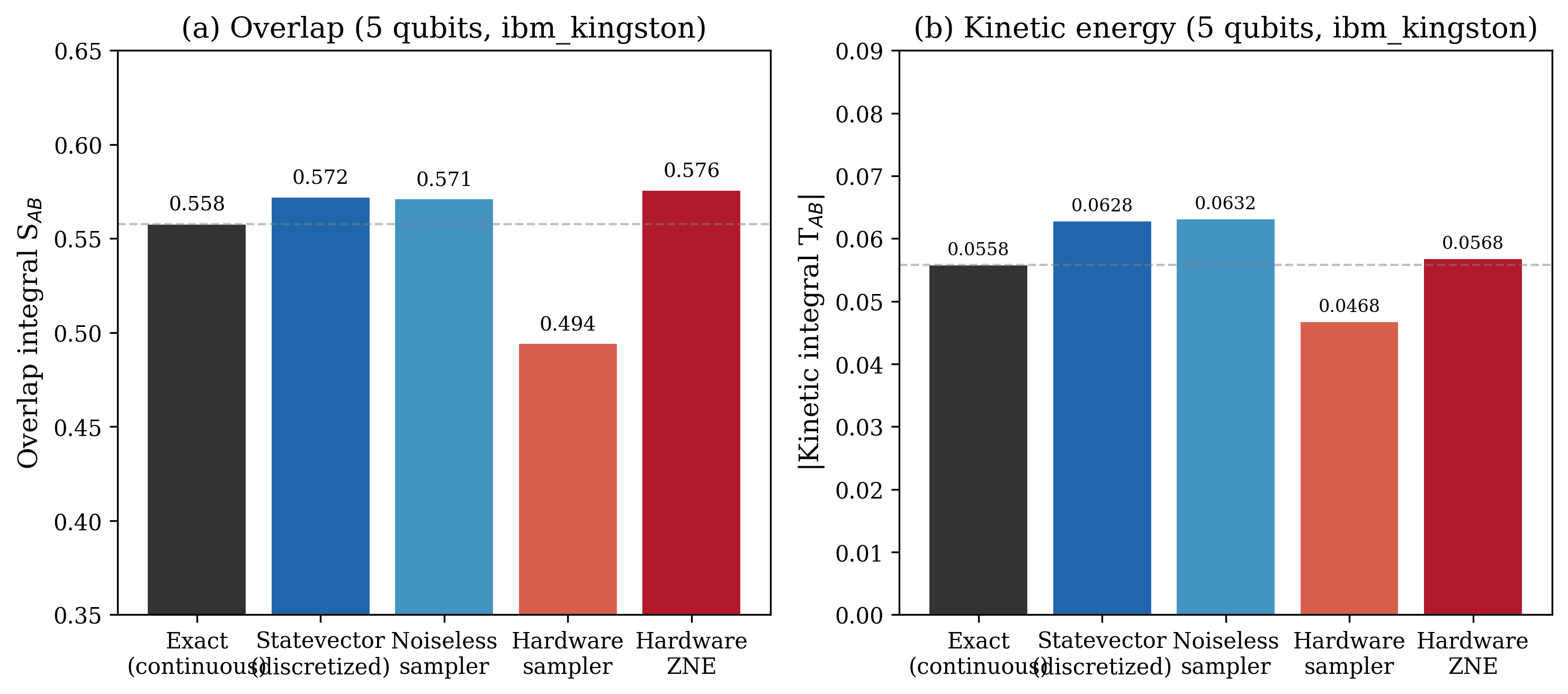}
\caption{\label{fig:hardware} Hardware validation at 5~qubits on ibm\_kingston. (a)~Overlap integral. (b)~Kinetic energy integral (absolute value). Dashed line: exact continuous value.}
\end{figure}

The best hardware result is the 5-qubit 1s overlap on ibm\_kingston with ZNE, yielding $S_{AB} = 0.576$ versus the statevector value of 0.572---a hardware-induced error of 0.67\% [Fig.~\ref{fig:hardware}; Table~\ref{tab:1d_integrals}]. The total error relative to the exact continuous integral is 3.2\%, dominated by the 2.5\% discretization error inherent to the 5-qubit grid. The transpiled circuit contains 104~CZ gates at depth 266 (Table~\ref{tab:circuits}).

\begin{figure}
\includegraphics[width=\columnwidth]{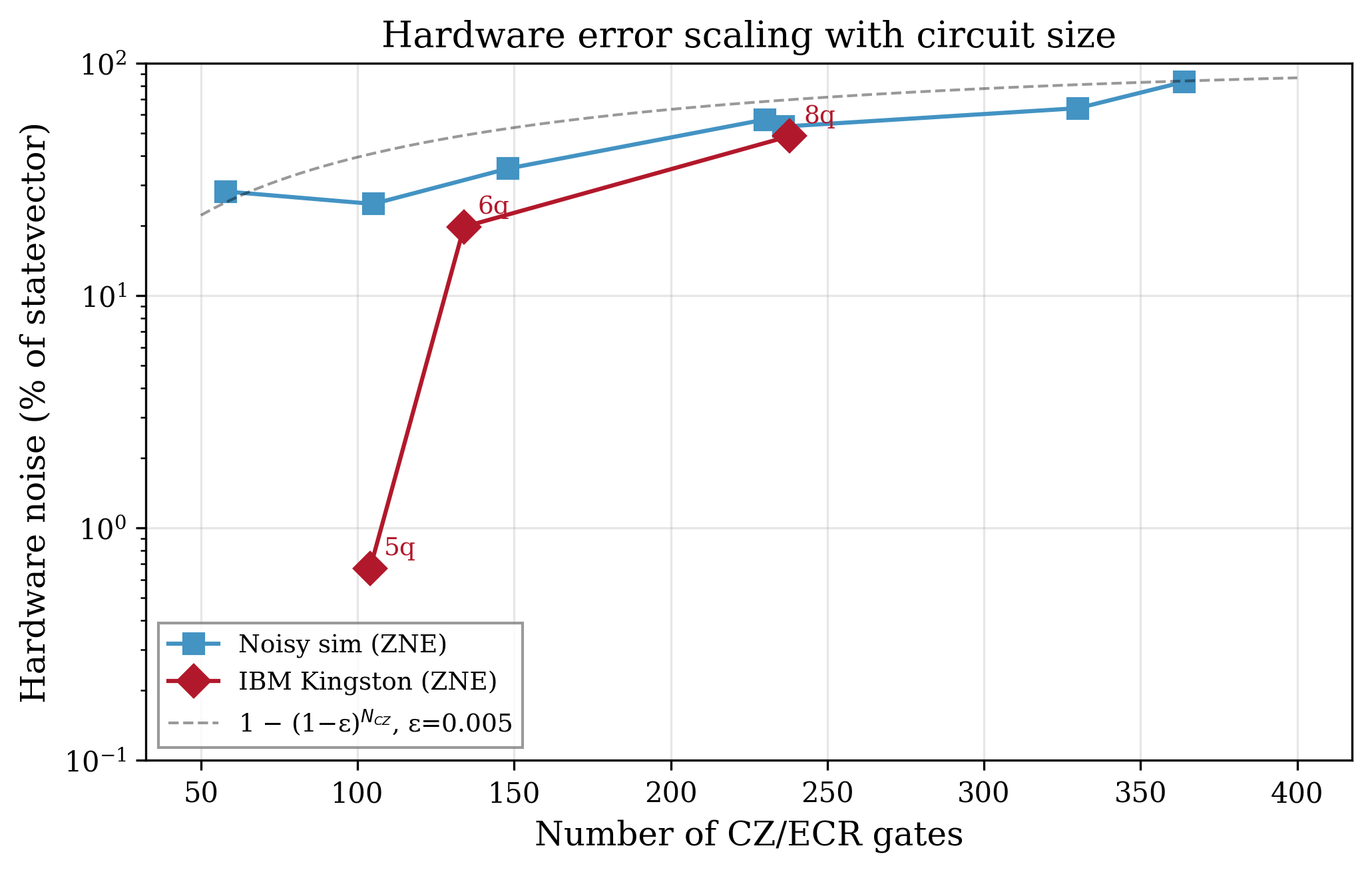}
\caption{\label{fig:error_scaling} Hardware error versus CZ gate count for the overlap integral. Dashed line: theoretical survival probability $(1-\varepsilon)^{N_\text{CZ}}$ with $\varepsilon = 0.5\%$.}
\end{figure}

At 6~qubits (134~CZ gates), the hardware error increases to 19.8\%, and at 8~qubits (238~CZ gates) it reaches 48.6\%. This degradation is consistent with the expected circuit survival probability: at a per-gate error rate $\varepsilon \approx 0.5\%$, the probability of error-free execution scales as $(1 - \varepsilon)^{N_\text{CZ}}$~\cite{Temme2017}, giving approximately 59\% survival at 104~gates and 30\% at 238~gates (Fig.~\ref{fig:error_scaling}).

\begin{table*}[t]
\caption{\label{tab:1d_integrals} One-dimensional one-electron integrals at H$_2$
equilibrium bond distance ($d = 1.4$~a.u., 16~a.u.\ simulation range).
``Exact'' is the analytical integral at the nearest discretized grid distance.
Disc.\ err.\ $= |\text{statevec}-\text{exact}|/|\text{exact}|$.
HW noise $= |\text{HW ZNE}-\text{statevec}|/|\text{statevec}|$.
Dashes indicate hardware data not obtained (circuit infeasible or signal below noise floor).}
\begin{ruledtabular}
\begin{tabular}{lcrrrrrl}
Integral & $n$ & Exact & Statevec. & Disc.\ err. & HW (ZNE) & HW noise & Backend \\
\hline
$S_{AB}$ (overlap) & 5 & 0.5578 & 0.5719 & 2.53\% & 0.5758 & 0.67\% & ibm\_kingston \\
 & 6 & 0.5578 & 0.5613 & 0.63\% & 0.4503 & 19.77\% & ibm\_kingston \\
 & 8 & 0.6005 & 0.6007 & 0.04\% & 0.3089 & 48.58\% & ibm\_kingston \\
 & 5 & 0.5578 & 0.5719 & 2.53\% & 0.5460 & 4.54\% & ibm\_marrakesh \\
$T_{AB}$ (kinetic) & 5 & $-0.0558$ & $-0.0628$ & 12.66\% & $-0.0568$ & 9.60\% & ibm\_kingston \\
 & 8 & $-0.0474$ & $-0.0475$ & 0.24\% & --- & --- & (sim only) \\
$V_{AB}$ (nuc.\ attr.) & 6 & --- & $-1.1153$ & --- & --- & --- & (sim only) \\
\end{tabular}
\end{ruledtabular}
\end{table*}

\subsubsection{Kinetic energy integral on hardware}

The kinetic energy integral was measured on ibm\_kingston at 5~qubits with ZNE, yielding $T_{AB} = -0.0568$ (Fig.~\ref{fig:hardware}) versus the statevector value $-0.0628$ (9.6\% hardware noise). The larger relative noise compared to the overlap arises from the small absolute value of the matrix element: the probability $P(0) = |T_{AB}|^2 \approx 0.004$ sits an order of magnitude below the hardware noise floor of approximately $1/2^5 \approx 0.03$ (the depolarization limit). The raw hardware sampler result is not reported for the kinetic energy, as the signal is indistinguishable from noise without error mitigation.

The kinetic energy circuit is nearly identical to the overlap circuit---one additional $Z$ gate per preparation---giving 107~CZ gates versus 104 for the overlap. The increased error is therefore attributable entirely to the signal-to-noise ratio, not to circuit complexity. This represents a general limitation of the compute/uncompute method for small matrix elements: any integral where the normalized overlap is small will produce $P(0)$ values that compete with the hardware noise floor. In the fault-tolerant setting, quantum amplitude estimation~\cite{Brassard2002} would amplify the signal quadratically, alleviating this limitation.

\subsubsection{Circuits beyond current hardware}

The 3D orthogonality circuit $\braket{1s|2s}$ in spherical coordinates requires 149--310~ECR gates at 4--8~qubits per coordinate (Table~\ref{tab:circuits}). While these circuits are more compact than the Cartesian 3D encodings, noisy simulation on the \texttt{fake\_sherbrooke} noise model shows that hardware noise overwhelms the signal at 5~qubits and above: the reported overlap of 0.18 at 5~qubits is dominated by noise against a true value of 0.019. The nuclear attraction circuit (8,486~ECR gates) and all 3D Cartesian circuits ($\geq$9,500~ECR gates) are validated exclusively in simulation.

The $O(n)$ scaling of circuit depth with qubit count implies that the method will become hardware-executable as gate fidelities improve, without requiring algorithmic changes. The threshold sensitivity analysis (Sec.~\ref{sec:entanglement}) provides a concrete resource reduction path: at truncation threshold $10^{-6}$, the 3D circuit resources would decrease by approximately a factor of 3.4 in bond dimension, with a corresponding reduction in transpiled gate count.

%% ============================================================
%% SECTION IV: DISCUSSION
%% ============================================================

\section{Discussion}
\label{sec:discussion}

\subsection{Entanglement as the organizing principle}

The results of this work establish the MPS bond dimension as a universal, quantitative criterion for the feasibility of amplitude encoding. Rather than a binary classification of ``efficient'' or ``intractable,'' the bond dimension defines a continuous cost metric that maps the full landscape of encoding complexity for atomic orbitals:

\begin{itemize}
\item \emph{1D, separable coordinates}: $\chi$ is constant (2--4 for analytical functions, 11 for $V \cdot \psi$). Circuit cost is $O(n)$ two-qubit gates. State preparation requires $O(n)$ classical operations with no grid sampling when analytical MPS constructions are available.

\item \emph{3D, spherical coordinates}: $\chi = 1$ across coordinate boundaries. Each coordinate is encoded independently using the 1D constructions. Total cost is the sum of three 1D preparations---the most efficient path.

\item \emph{3D, Cartesian coordinates}: $\chi_{\max}$ saturates at $\sim$138 (threshold $10^{-12}$) or $\sim$39 (threshold $10^{-6}$) for hydrogen 1s. The cross-coordinate entanglement $\chi(x|y)$ also saturates, reaching 41 at 11~qubits per coordinate and remaining unchanged at 12. Circuit cost is $O(n)$ in depth but with a large per-gate constant determined by $\lceil\log_2(\chi_{\max})\rceil$ bond qubits.

\item \emph{Two-electron Coulomb kernel (1D)}: full Schmidt rank equal to the grid size. No efficient encoding exists in one dimension.
\end{itemize}

Each transition in this hierarchy has a physical origin. The 1D separability arises from the factorization of exponentials over binary digits. The spherical separability comes from the product structure $\psi(r,\theta,\varphi) = R(r) \cdot \Theta(\theta) \cdot \Phi(\varphi)$. The Cartesian entanglement arises from the nonlinear coupling $r = \sqrt{x^2 + y^2 + z^2}$, which mixes coordinate registers but does so in a bounded way---the radial structure has finite complexity that saturates once the grid resolves it. The two-electron full rank arises from the $1/|x_1 - x_2|$ kernel, which is irreducibly coupled across the two electron coordinates in one dimension.

\subsection{Cartesian encoding: bounded but expensive}

The saturation of bond dimension in Cartesian coordinates is perhaps the most unexpected finding of this work. An initial analysis at small qubit counts suggested exponential growth ($\chi \sim 2^n$), but systematic measurements up to 12~qubits per coordinate (36 total qubits, $4096^3 \approx 7 \times 10^{10}$ grid points) reveal clear deceleration and convergence. The successive increments $\Delta\chi_{\max}$ peak at $n = 6$ and decrease monotonically thereafter. The cross-coordinate entanglement $\chi(x|y)$ stops growing entirely between $n = 11$ and $n = 12$.

This saturation has practical implications. At threshold $10^{-12}$, the asymptotic $\chi_{\max} \approx 138$ requires 8~bond qubits, meaning each MPS gate acts on 9~qubits. At threshold $10^{-6}$, $\chi_{\max} \approx 39$ requires only 6~bond qubits (7-qubit gates). The transpiled two-qubit gate count per MPS tensor is correspondingly smaller, though still substantial---the 3D 1s state preparation at 3~qubits per coordinate already produces $\sim$9,500 ECR gates (Table~\ref{tab:circuits}).

Cartesian encoding is therefore not fundamentally forbidden but quantitatively expensive. Spherical coordinates remain strongly preferred for single-center quantities. However, for multi-center problems where Cartesian grids may be unavoidable---or where coordinate transformations introduce their own complexity---the bounded bond dimension means that MPS-based preparation is feasible in principle. The cost is a large but finite constant, independent of the target grid resolution.

The SVD truncation threshold provides a practical resource knob: reducing from $10^{-12}$ to $10^{-6}$ cuts the bond dimension by a factor of 3.4 with no measurable impact on the computed integrals. For applications where chemical accuracy ($\sim 10^{-6}$~Hartree) is sufficient, the lower threshold offers a direct path to smaller circuits.

\subsection{Classical preprocessing and scalability}

The amplitude encoding pipeline consists of two stages: classical preprocessing (constructing the MPS tensors) and quantum execution (running the circuit). The total cost is the sum of both.

For 1D orbital functions with analytical MPS constructions (1s, 2s, derivatives), the classical cost is $O(n)$---the tensor elements are computed from the orbital parameters and grid spacing using constant-size matrices. No function evaluations on the grid are needed. This is the strongest form of efficiency: the entire pipeline, classical and quantum, scales as $O(n)$.

For functions without analytical MPS ($V \cdot \psi$ in 1D, all 3D Cartesian encodings), the classical preprocessing requires evaluating the function on the grid to perform the SVD. The brute-force cost is $O(2^n)$ in 1D and $O(2^{3n})$ in 3D---exponential in the number of qubits. This does not negate the quantum advantage in principle, since the SVD is performed once and the resulting circuit can be executed repeatedly for different integral evaluations. However, it does limit the practical qubit counts for which the MPS can be computed classically.

TT-cross interpolation~\cite{Oseledets2010} provides a scalable alternative. This algorithm constructs an approximate MPS from $O(n \cdot \chi^2)$ adaptively chosen function evaluations, exploiting the low-rank structure without requiring the full grid. For the 1D $V \cdot \psi$ function ($\chi = 11$), TT-cross would require $\sim$1,200 evaluations at $n = 10$ versus 1,024 for brute force---comparable at this scale but with fundamentally better asymptotic scaling. For the 3D Cartesian case ($\chi_{\max} \sim 138$), the savings become significant at larger qubit counts: $O(n \cdot 138^2) \approx O(19{,}000 \cdot n)$ evaluations versus $O(2^{3n})$. At $n = 10$ qubits per coordinate, this is $\sim$190,000 evaluations versus $\sim 10^9$---a reduction of four orders of magnitude.

The complete scalability picture is therefore: analytical MPS where available ($O(n)$ everything), TT-cross where not ($O(n \cdot \chi^2)$ classical, $O(n)$ quantum), with the bond dimension $\chi$ as the governing parameter in both cases.

\subsection{Limitations and signal-to-noise}

Several limitations of the current approach merit discussion.

\emph{Hardware noise and circuit depth.} The compute/uncompute method requires executing the full state preparation circuit twice ($U_B$ followed by $U_A^\dagger$), doubling the gate count relative to a single preparation. At 5~qubits with $\sim$105~CZ gates, the overlap integral is measured with 0.67\% hardware noise. The kinetic energy, using a nearly identical circuit, suffers 9.6\% noise---not because of circuit complexity but because of signal weakness. Scaling to larger qubit counts or higher bond dimensions will require error rates substantially below current levels, or the use of quantum error correction.

\emph{Signal-to-noise for small matrix elements.} The compute/uncompute method returns $P(0) = |\braket{\psi_A|\psi_B}|^2$. When the matrix element is small, $P(0)$ approaches the depolarization floor of $\sim 1/2^n$, and the signal becomes indistinguishable from noise. This is a fundamental limitation of the method, not specific to our implementation. The kinetic energy integral ($P(0) \approx 0.004$ at 5~qubits, versus a noise floor of $\sim$0.03) illustrates this directly. Quantum amplitude estimation~\cite{Brassard2002} would improve the scaling from $O(1/\varepsilon^2)$ shots to $O(1/\varepsilon)$, providing a quadratic advantage in signal extraction that would alleviate this limitation in the fault-tolerant setting.

\emph{Divergence of 1D Coulomb integrals.} The nuclear attraction integral $\int \psi_A^*(x) \cdot (1/|x - R_C|) \cdot \psi_B(x)\, dx$ diverges in one dimension for all orbital configurations, because the $1/|x|$ singularity is not integrable in 1D. The discretized computation yields a well-defined but grid-dependent value that does not converge with increasing resolution. This limitation is specific to the 1D model; in three dimensions, the volume element $r^2 dr$ renders the integral finite, and the $V \cdot \psi$ encoding methodology developed here applies directly.

\emph{Classical preprocessing for 3D.} The numerical MPS decomposition for 3D Cartesian encodings currently requires evaluating the target function on all $2^{3n}$ grid points, limiting practical demonstrations to $\sim$7~qubits per coordinate. TT-cross interpolation (Sec.~\ref{sec:discussion}.C) provides the algorithmic solution but has not yet been implemented in our pipeline.

\emph{Qubit connectivity.} The MPS circuit applies sequential gates that each act on one physical qubit and all bond qubits. On hardware with limited connectivity, SWAP gates are needed to bring non-adjacent qubits together, inflating the two-qubit gate count. The transpiled circuits reported in this work include this overhead. Hardware architectures with higher connectivity or native multi-qubit gates would reduce this cost.

\subsection{Path to multi-center integrals and future work}

The methodology developed here---MPS-based amplitude encoding, compute/uncompute overlap extraction, and $V \cdot \psi$ product encoding for potential operators---provides the building blocks for a complete quantum chemistry integral evaluation framework. Several extensions are natural next steps.

\emph{3D kinetic energy and nuclear attraction integrals.} The $V \cdot \psi$ encoding method validated in 1D applies directly to three dimensions, where the Coulomb integral is finite. The key open question is the bond dimension of $V(r) \cdot \psi(r,\theta,\varphi)$ in spherical coordinates: if it remains bounded (as in the 1D case, $\chi = 11$), the nuclear attraction matrix element can be computed using the same overlap machinery. The kinetic energy integral in 3D requires the Laplacian in the chosen coordinate system, which decomposes into radial and angular terms amenable to the derivative encoding developed here. These are the most immediate extensions.

\emph{Multi-center orbital representation.} For integrals involving orbitals centered at different positions, the orbitals must be represented in a common coordinate system. In spherical coordinates, the Barnett-Coulson single-center expansion~\cite{BarnettCoulson1951} expresses an off-center orbital as a series in angular momentum channels about a common origin: each term is separable in spherical coordinates and amenable to efficient MPS encoding. The angular sum converges for well-separated centers, and the quantum computer handles the radial complexity while the angular coupling is treated as a classical summation. However, for close centers the expansion may converge slowly, and the bond dimension of the radial expansion functions $g_l(r)$ requires investigation.

\emph{Two-electron integrals.} The full Schmidt rank of the 1D Coulomb kernel precludes efficient encoding in one dimension. In three dimensions, the multipole expansion of $1/|r_1 - r_2|$ combined with the addition theorem for spherical harmonics decomposes the two-body operator into a sum of separable terms, each factorizing into functions of the individual electron coordinates. This provides a path to efficient two-electron integral evaluation that stays within the separable regime established here.

\emph{Fault-tolerant execution.} The 3D Cartesian circuits ($\geq$9,500~ECR gates) and the nuclear attraction circuit (8,486~ECR gates) are infeasible on current noisy hardware but well within the capabilities projected for early fault-tolerant quantum computers. The bounded bond dimension ensures that the circuit depth scales as $O(n)$ regardless of the target precision, meaning that hardware improvements translate directly into larger accessible systems without algorithmic changes. The threshold sensitivity analysis provides a concrete resource-accuracy trade-off for circuit design in the fault-tolerant setting.

\emph{Extension to heavier atoms and molecules.} The sensitivity of bond dimension to the orbital decay constant $\zeta$ (Sec.~\ref{sec:entanglement}) suggests that Cartesian encoding costs decrease for heavier atoms, where orbital exponents are larger and the functions are more localized. A systematic study across the periodic table would map the encoding cost as a function of atomic number, providing a practical guide for quantum chemistry applications.

%% ============================================================
%% SECTION V: CONCLUSION
%% ============================================================

\section{Conclusion}
\label{sec:conclusion}

We have presented a systematic study of amplitude encoding of Slater-type orbitals on quantum computers using matrix product states. The approach is constructive---circuits are built directly from the MPS decomposition of the target function, not through variational optimization---and the bond dimension serves as a measurable, quantitative predictor of circuit cost.

For one-dimensional orbital functions of the form $p_d(x) \cdot e^{-\zeta x}$, we derived analytical MPS constructions with constant bond dimension $\chi = d + 1$, where $d$ is the polynomial degree. These constructions require $O(n)$ classical and quantum operations, with no sampling of the $2^n$-point grid. The 1s orbital ($\chi = 2$), 2s STO ($\chi = 2$), and hydrogen 2s with Jacobian ($\chi = 3$) all admit closed-form transfer matrices derivable from the orbital parameters alone.

Using these encodings, we demonstrated in one dimension the computation of three types of one-electron integrals---overlap, kinetic energy, and nuclear attraction---and validated the overlap and kinetic energy on IBM Heron quantum processors at 5~qubits, achieving 0.67\% hardware noise on the overlap integral with Zero-Noise Extrapolation. The nuclear attraction integral was validated in simulation using a product-state encoding of $V(x) \cdot \psi(x)$ with saturating bond dimension $\chi = 11$. In three dimensions, we computed multi-center overlap integrals between 1s and 2s orbitals in Cartesian coordinates, achieving 0.02\% discretization error at 6~qubits per coordinate (18 total qubits) in statevector simulation. The extension of kinetic energy and nuclear attraction integrals to three dimensions, where the Coulomb potential is integrable, is left for future work.

A systematic entanglement analysis across coordinate systems revealed a richer landscape than previously understood. In three-dimensional Cartesian coordinates, both the maximum bond dimension and the cross-coordinate entanglement saturate with increasing grid resolution, reaching asymptotic values of $\sim$138 and $\sim$41 respectively for the hydrogen 1s orbital. This saturation---confirmed up to 12~qubits per coordinate ($4096^3$ grid points)---establishes that Cartesian amplitude encoding of STOs has bounded complexity: expensive but not intractable. The SVD truncation threshold provides a practical tuning parameter, reducing the bond dimension from 138 to 39 at threshold $10^{-6}$ with negligible impact on integral accuracy. Spherical coordinates remain the most efficient encoding ($\chi = 1$ across coordinate boundaries), but Cartesian coordinates are not fundamentally forbidden.

These results establish amplitude encoding via MPS as a viable path toward exact STO basis sets on quantum computers. The immediate next steps are three-dimensional kinetic energy and nuclear attraction integrals (where the Coulomb potential is integrable), multi-center orbital representation via single-center angular momentum expansions, and two-electron integrals via the multipole decomposition of the Coulomb operator. Each of these extensions builds directly on the encoding and integral evaluation machinery developed here, staying within the separable coordinate framework where efficient state preparation is guaranteed.

% %% ============================================================
% %% ACKNOWLEDGMENTS
% %% ============================================================

\begin{acknowledgments}
The source code for all MPS constructions, integral evaluations, and entanglement analyses is available at \url{https://github.com/sorin-bolos/papers/tree/main/AmplitudeEncodingOfSlaterTypeOrbitals}.
\end{acknowledgments}

%% ============================================================
%% REFERENCES
%% ============================================================


\begin{thebibliography}{0}%
\makeatletter
\providecommand \@ifxundefined [1]{%
 \@ifx{#1\undefined}
}%
\providecommand \@ifnum [1]{%
 \ifnum #1\expandafter \@firstoftwo
 \else \expandafter \@secondoftwo
 \fi
}%
\providecommand \@ifx [1]{%
 \ifx #1\expandafter \@firstoftwo
 \else \expandafter \@secondoftwo
 \fi
}%
\providecommand \natexlab [1]{#1}%
\providecommand \enquote  [1]{``#1''}%
\providecommand \bibnamefont  [1]{#1}%
\providecommand \bibfnamefont [1]{#1}%
\providecommand \citenamefont [1]{#1}%
\providecommand \href@noop [0]{\@secondoftwo}%
\providecommand \href [0]{\begingroup \@sanitize@url \@href}%
\providecommand \@href[1]{\@@startlink{#1}\@@href}%
\providecommand \@@href[1]{\endgroup#1\@@endlink}%
\providecommand \@sanitize@url [0]{\catcode `\\12\catcode `\$12\catcode
  `\&12\catcode `\#12\catcode `\^12\catcode `\_12\catcode `\%12\relax}%
\providecommand \@@startlink[1]{}%
\providecommand \@@endlink[0]{}%
\providecommand \url  [0]{\begingroup\@sanitize@url \@url }%
\providecommand \@url [1]{\endgroup\@href {#1}{\urlprefix }}%
\providecommand \urlprefix  [0]{URL }%
\providecommand \Eprint [0]{\href }%
\providecommand \doibase [0]{https://doi.org/}%
\providecommand \selectlanguage [0]{\@gobble}%
\providecommand \bibinfo  [0]{\@secondoftwo}%
\providecommand \bibfield  [0]{\@secondoftwo}%
\providecommand \translation [1]{[#1]}%
\providecommand \BibitemOpen [0]{}%
\providecommand \bibitemStop [0]{}%
\providecommand \bibitemNoStop [0]{.\EOS\space}%
\providecommand \EOS [0]{\spacefactor3000\relax}%
\providecommand \BibitemShut  [1]{\csname bibitem#1\endcsname}%
\let\auto@bib@innerbib\@empty
%</preamble>
\end{thebibliography}%


\begin{thebibliography}{99}

\bibitem{Slater1930} J.~C.~Slater, Phys.\ Rev.\ \textbf{36}, 57 (1930).

\bibitem{HehreStewartPople1969} W.~J.~Hehre, R.~F.~Stewart, and J.~A.~Pople, J.\ Chem.\ Phys.\ \textbf{51}, 2657 (1969).

\bibitem{SzaboOstlund} A.~Szabo and N.~S.~Ostlund, \textit{Modern Quantum Chemistry: Introduction to Advanced Electronic Structure Theory} (Dover, New York, 1996).

\bibitem{Shende2006} V.~V.~Shende, S.~S.~Bullock, and I.~L.~Markov, IEEE Trans.\ CAD \textbf{25}, 1000 (2006).

\bibitem{Peruzzo2014} A.~Peruzzo \textit{et al.}, Nat.\ Commun.\ \textbf{5}, 4213 (2014).

\bibitem{Babbush2018} R.~Babbush \textit{et al.}, Phys.\ Rev.\ X \textbf{8}, 041015 (2018).

\bibitem{Haner2018} T.~H\"aner, M.~Roetteler, and K.~M.~Svore, Quantum Sci.\ Technol.\ \textbf{3}, 035005 (2018).

\bibitem{McClean2018} J.~R.~McClean, S.~Boixo, V.~N.~Smelyanskiy, R.~Babbush, and H.~Neven, Nat.\ Commun.\ \textbf{9}, 4812 (2018).

\bibitem{Schon2005} C.~Sch\"on, E.~Solano, F.~Verstraete, J.~I.~Cirac, and M.~M.~Wolf, Phys.\ Rev.\ Lett.\ \textbf{95}, 110503 (2005).

\bibitem{Ran2020} S.-J.~Ran, Phys.\ Rev.\ A \textbf{101}, 032310 (2020).

\bibitem{NielsenChuang} M.~A.~Nielsen and I.~L.~Chuang, \textit{Quantum Computation and Quantum Information}, 10th Anniversary ed.\ (Cambridge University Press, Cambridge, 2010).

\bibitem{Brassard2002} G.~Brassard, P.~H\o yer, M.~Mosca, and A.~Tapp, Contemp.\ Math.\ \textbf{305}, 53 (2002).

\bibitem{Schollwock2011} U.~Schollw\"ock, Ann.\ Phys.\ \textbf{326}, 96 (2011).

\bibitem{Oseledets2010} I.~V.~Oseledets and E.~E.~Tyrtyshnikov, SIAM J.\ Sci.\ Comput.\ \textbf{33}, 2295 (2011).

\bibitem{IBMQuantum} IBM Quantum Platform, Processor types, https://quantum.cloud.ibm.com/docs/en/guides/processor-types (accessed April 2026).

\bibitem{Qiskit} Qiskit contributors, ``Qiskit: An open-source framework for quantum computing'' (2023), doi:10.5281/zenodo.2573505.

\bibitem{Temme2017} K.~Temme, S.~Bravyi, and J.~M.~Gambetta, Phys.\ Rev.\ Lett.\ \textbf{119}, 180509 (2017).

\bibitem{Li2017} Y.~Li and S.~C.~Benjamin, Phys.\ Rev.\ X \textbf{7}, 021050 (2017).

\bibitem{BarnettCoulson1951} M.~P.~Barnett and C.~A.~Coulson, Phil.\ Trans.\ R.\ Soc.\ Lond.\ A \textbf{243}, 221 (1951).

\end{thebibliography}
\end{document}